\documentclass[twocolumn,showpacs,preprintnumbers,amsmath,amssymb,prb,floatfix]{revtex4}

\usepackage{graphicx}
\usepackage{dcolumn}
\usepackage{bm}
\usepackage{epstopdf}  
 
\begin{document}
\bibliographystyle{prsty}    
 

\title{Nuclear Tunneling and Dynamical Jahn-Teller Effect in Graphene with  Vacancy} 

\author{Z. S. Popovi\'c*}
\author{B. R. K. Nanda$^\dagger$}
\author{S. Satpathy}

\affiliation{
Department of Physics $\&$ Astronomy, University of Missouri,  
Columbia, MO 65211}

\date{\today}     

\begin{abstract}
We show that the substitutional vacancy in graphene forms a dynamical Jahn-Teller center. The adiabatic potential surface resulting from the  electron-lattice coupling was computed using density-functional methods and subsequently the Schr\"odinger equation was solved for the nuclear motion. Our calculations show a large tunneling splitting $3 \Gamma$ of about 86 cm$^{-1}$.
The effect results in  a large delocalization of the carbon nuclear wave functions around the vacancy leading to a significant broadening of the Jahn-Teller active $sp^2\sigma$ electron states. The tunneling splitting should be observable in electron paramagnetic resonance and two-photon resonance scattering experiments.

\end{abstract}
\maketitle

In spite of its deceptively simple honeycomb lattice structure,   graphene has quickly become a new paradigm for testing a variety of ideas in condensed matter physics. 
The   much celebrated linear band structure of graphene\cite{Wallace} leads to a host of unusual behaviors such as Klein tunneling, chiral electrons, minimum conductivity, negative refraction, half-integer quantum Hall effect, and new features in the Kondo and RKKY interactions leading to quantum criticality.\cite{graphene-Novoselov-2004, Review1, Review2, Review3}  
 Vacancies in the carbon based systems have been of considerable interest for quite some time now, especially in the context of magnetism without magnetic atoms.\cite{carbon-vacancy-magnetism-review, Butz, Garcia, nieminen, Nair}
Quite remarkably, it has been shown that a vacancy introduces a quasi-localized midgap state in the $\pi$ bands with $\sim 1/ r$ decay on account of the particle-hole symmetry.\cite{castro-zero-mode, Nanda-Vacancy,Ugeda} An interesting consequence of this is the partial occupation of the vacancy-induced $\sigma$-band states, which leads then to a Jahn-Teller (JT) distortion. 
The JT distortion could be static or dynamic. In the latter, the potential barrier between the different equivalent minima in the nuclear configuration space is small enough that the nuclei tunnel between the various minima leading to several interesting effects, while in the static JT effect, the nuclei are stuck to one minima or the other.
In this Letter, we show that the vacancy forms a dynamical JT center in graphene owing to the small quantum mechanical barrier for nuclear tunneling.


 %
\begin{figure}
\includegraphics[angle=0,width=0.85   \linewidth]{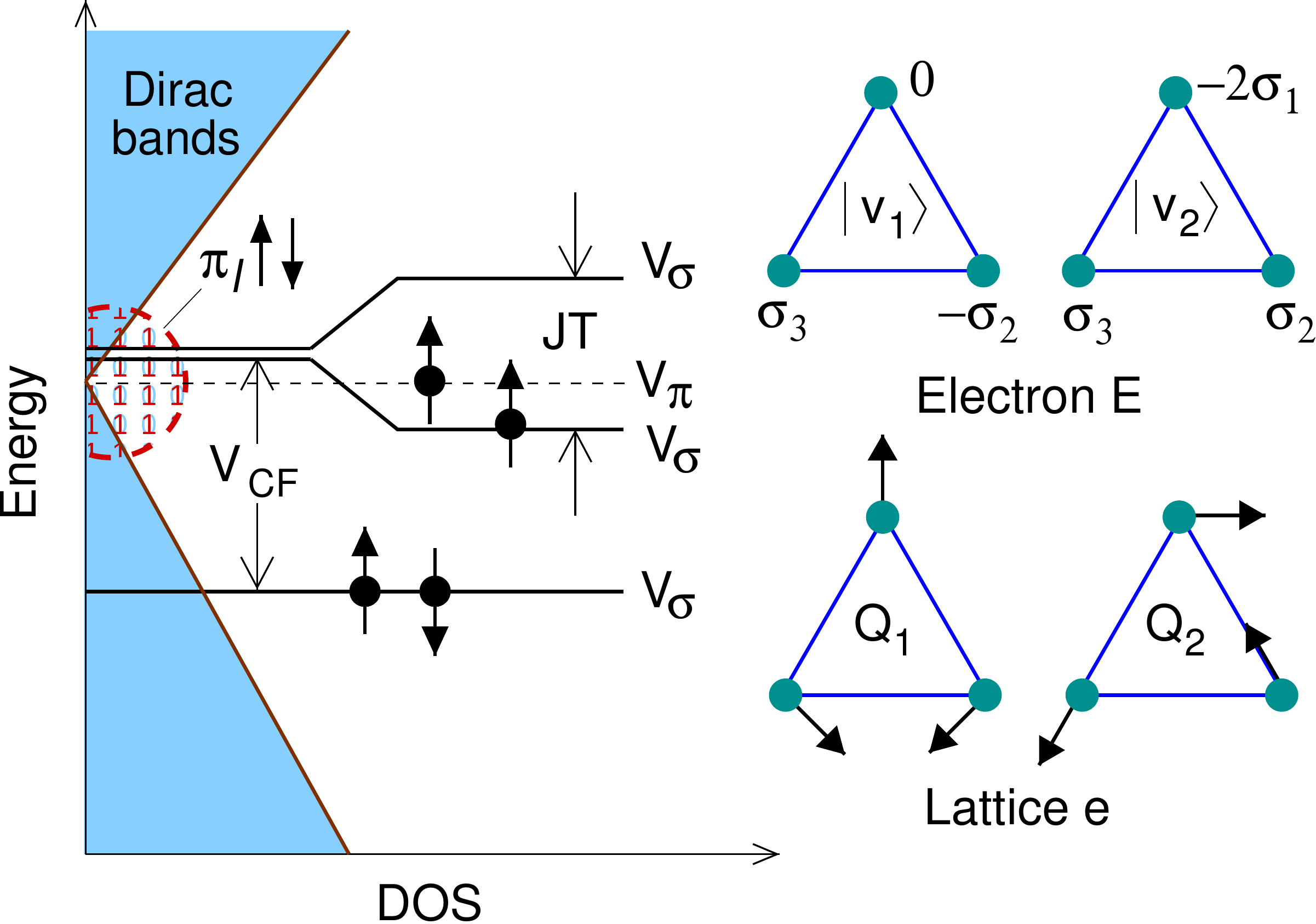}
\caption{ (Color online) Vacancy induced $\sigma$ and $\pi$ electron states (V$_\sigma$ and V$_\pi$)  with the occupied states shown by dots with arrows ({\it left}).  
The nominal $2 \mu_B$ ($S = 1$) magnetic moment due to the localized states is reduced substantially due to the anti-ferromagnetic spin polarization of the band states, indicated by  $\pi_l \uparrow \downarrow$,  in the local neighborhood of the vacancy. 
{\it Right} part shows the JT active electron states, $|v_1\rangle$ and $|v_2\rangle$, and the vibrational modes of the carbon triangle that they couple to. $\sigma_i$ denotes the dangling $sp^2\sigma$ bond orbital on a carbon atom adjacent to the vacancy. 
}
\label{schematic}
\end{figure}
\begin{figure}[!htb]
\includegraphics[angle=0,width=0.65\linewidth]{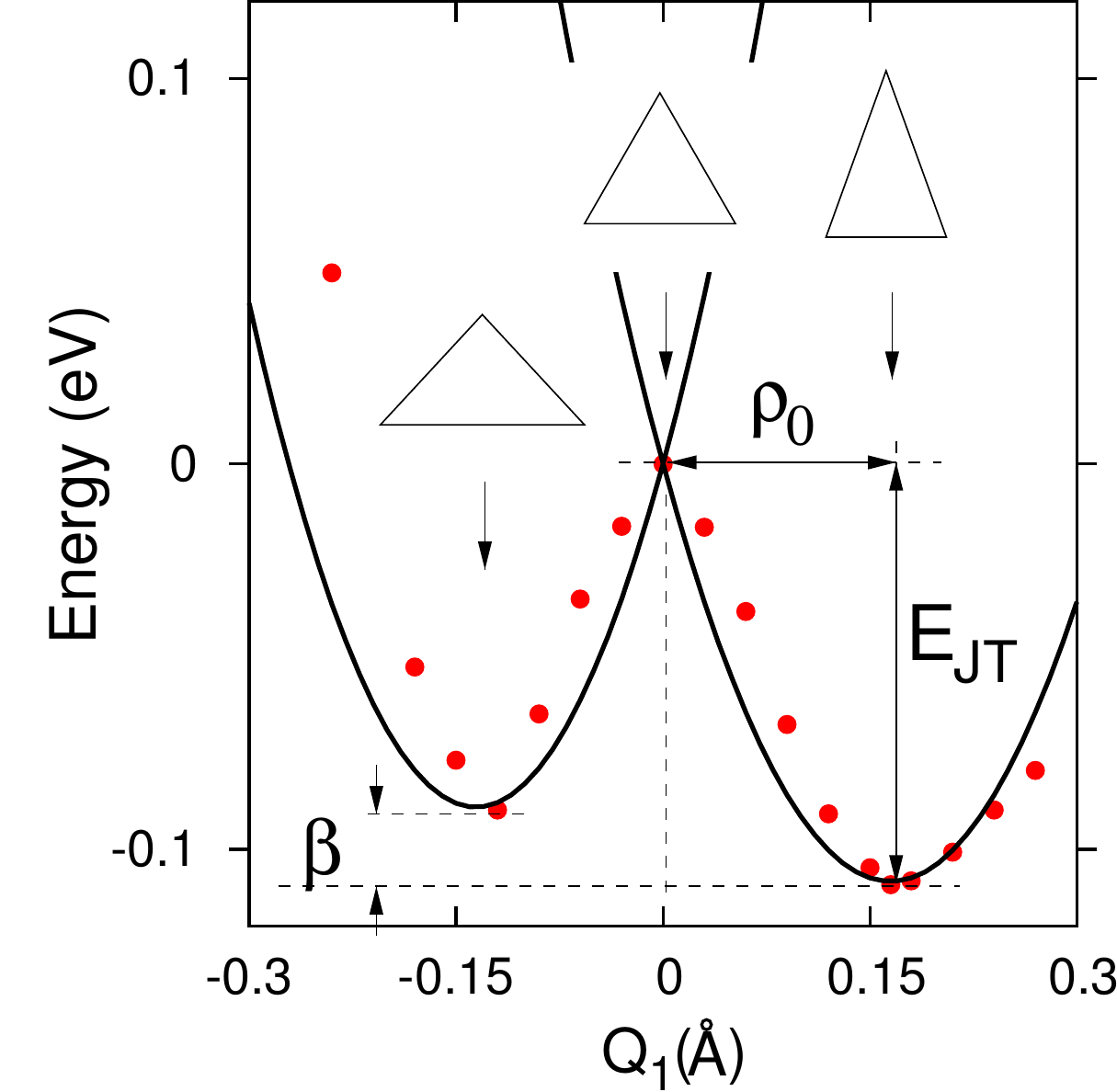}      
\includegraphics[angle=0,width=0.65\linewidth]{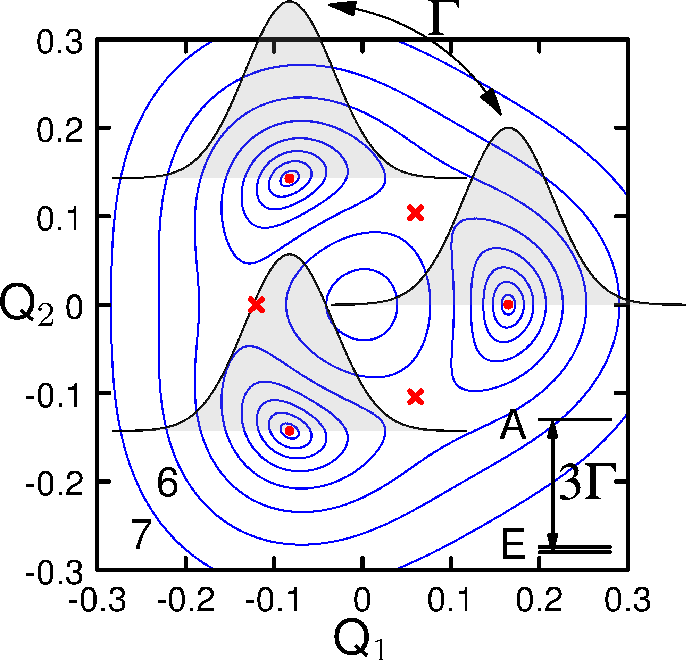}
\caption{(Color online)  
Total energy as a function of the vibronic distortion $Q_1$  computed from the DFT (red dots) and  fitted to the adiabatic energy $E_-$ in Eq. (\ref{APES})(full line) {\it(Top)}. 
The triangles indicate the configurations at the three extrema. {\it Bottom} figure shows the corresponding energy contours in the $Q_1-Q_2$ plane, with the three equivalent minima (dots) separated by the tunneling barriers (crosses). The contour values are: $-0.11 + 0.001 \times (2^n)$ in units of eV, where $n = 0, 1, ..., 7 $ labels the contours and $\Gamma$ denotes the  nuclear hopping integral in the tight binding description. 
}
\label{LAPW}
\end{figure}

 
Density-functional calculations\cite{Nanda-Vacancy} show that the vacancy introduces four electrons into the graphene bands as illustrated in Fig. \ref{schematic}. 
The JT effect comes from the partial occupation of the  doubly-degenerate $sp^2\sigma$ dangling bond states on the carbon triangle surrounding the vacancy and their coupling to the two vibrational modes of the triangle, given by the 
$E\otimes e$ JT Hamiltonian\cite{Bersuker2006,Tunneling-PRL2010} 
\begin{eqnarray}
\label{hamiltonian}
{\cal H} &=&-\frac{\hbar^2}{2M}\left(
\frac{\partial^2}{\partial Q_1^2}+
\frac{\partial^2}{\partial Q_2^2}
\right)
+
\frac{1}{2}K\left(Q_1^2+Q_2^2\right)
\\
&+& g  \left(Q_1\hat\tau_z+Q_2\hat\tau_x\right)
+G\left[\left(Q_1^2-Q_2^2\right)\hat\tau_z+2Q_1 Q_2\hat\tau_x\right],
\nonumber
\end{eqnarray}
where the terms are the nuclear kinetic energy, the elastic energy, and the linear and the quadratic JT coupling terms. 
Here the pseudospin $\vec \tau$ describes the two JT active, doubly-degenerate electronic states $|v_1\rangle$ and $|v_2\rangle$  originating from the three $sp^2\sigma$ dangling bonds  on the carbon triangle:
$|v_0\rangle=( \sigma_1 + \sigma_2 + \sigma_3)/\sqrt 3$,
$|v_1\rangle  =(     - \sigma_2 + \sigma_3)/\sqrt 2$,
$|v_2\rangle  =(2\sigma_1 - \sigma_2 - \sigma_3)/\sqrt 6$,
with
energies  $E_0=-2t$ and $E_{1,2}=t$ and symmetries $A_1$ and $E$, respectively,
with the $-t$ being the $\sigma$-electron hopping between the neighboring sites on the triangle, and $|v_1\rangle$ transforms like $x$ and $|v_2\rangle$ like $y$. 
On the other hand, the $p_z$ orbitals, responsible for the  linear `$\pi$' Dirac bands,  
introduce the quasi-localized midgap state, which  becomes singly-occupied due to Hund's coupling, leaving a lone electron to occupy the $\sigma$-derived doubly-degenerate $E$ state. 
This explains the the relative positions of the vacancy states shown in Fig. \ref{schematic}.
Turning now to the three vibrational modes of the triangle:
$\vert Q_0\rangle$ $\ =\ $ $(0,2,  \sqrt3,-1,    -\sqrt3,-1)/\sqrt{12}$,
$\vert Q_1\rangle$ $\ =\ $ $(0,2, -\sqrt3,-1,   \sqrt3,-1)/\sqrt{12}$,
$\vert Q_2\rangle$ $\ =\ $ $(2,0, -1,$ $ \sqrt3,   -1, -\sqrt3 )/ \sqrt{12}$,
\cite{Bersuker2006}
$Q_0$ is the stretching mode and the doubly-degenerate  $Q_1$ and $Q_2$ modes are JT active, splitting the upper two V$\sigma$ bands as shown in Fig. \ref{schematic}. 
The parameters in the Hamiltonian are the carbon mass $M$, the elastic energy $K$, and the linear and quadratic JT coupling 
parameters $g$ and $G$, respectively. 
Diagonalization of the potential terms in Eq. \ref{schematic} leads to the well-known 
adiabatic potential surface (APS) for the nuclear motion 
%
\begin{equation}
\label{APES}
E_\pm=\frac{1}{2}K\rho^2 
\pm\rho\sqrt{g^2+G^2\rho^2+2 g G\rho\cos(3\phi)},
\end{equation}
where
$\rho=\sqrt{Q_1^2+Q_2^2}$ and $\phi=\tan^{-1} (Q_2/Q_1)$ are the polar coordinates  and $E_\pm$ denote the two potential sheets. 
Without the quadratic coupling ($G=0$), one gets the Mexican hat APS, while with it we have three minima in the $(Q_1, Q_2)$ space (Fig. \ref{LAPW}).
The  electronic eigenfunction for the lower sheet is\cite{Bersuker2006}
\begin{equation}
|\psi^e \rangle = [\sin (\phi/2) |v_1 \rangle + \cos (\phi/2) |v_2 \rangle] \times e^{i\phi/2},
\label{Psi-electronic}
\end{equation}
 where the phase factor assures single-valuedness as one moves around the origin and leads to a Berry phase.

 In order to study the APS, we have computed the total energy as a function of the vibronic coordinates using the spin-polarized density functional all-electron linear augmented plane-waves (LAPW) method \cite{Wien-code} and the gradient approximation (GGA) for the exchange-correlation functional.\cite{Perdew-96} 
 We used a 32-atom supercell with a single vacancy and obtained a fully relaxed structure, which yielded a planar structure with an isosceles  triangle for the carbon atoms surrounding the vacancy with two long bonds (2.66 \AA)   and one short bond (2.41 \AA).  
This is equivalent to the distortion: Q$_0$ = 0.08 \AA, Q$_1$ = 0.166 \AA, and Q$_2$ = 0.
We then took a series of structures with varying distortions, Q$_1$ and Q$_2$,  and in each case optimized the rest of the carbon atoms in the supercell. We note that while the literature is divided regarding whether the relaxed structure with a vacancy is planar or non-planar, the three-fold symmetry of the adiabatic potential surface occurs in either case, being tied to the symmetry of the honeycomb lattice itself. 
The calculated energies are shown in  Fig. \ref{LAPW}, which yields  
the JT distortion radius $\rho_0=0.165$ \AA, the JT stabilization energy $E_{\text{JT}}$= 110 \ meV
and the tunneling barrier height $\beta$=19 \ meV. Comparison of these results with Eq. (\ref{APES}) yields the stiffness constant $K = 9.3\ eV$/ \AA$^2$ and the linear and the quadratic JT parameters $g = 1.46\ eV/$ \AA\  and $G = 0.38 \ eV/$ \AA$^2$, respectively.
For the case of LaMnO$_3$,  a well-known system with a strong JT interaction, while the $K$ and $g$ are about the same, the warping parameter $G = 2.0 \ eV/$ \AA$^2$ is significantly large,\cite{Popovic-JT} 
which results in a static JT effect with the nuclei stuck to one potential minimum.
In contrast, the weaker warping term $G$ in graphene leads to a small barrier height for nuclear tunneling and consequently to the dynamic JT effect, where the nuclei tunnel between the three minima in the APS. Since the phonon frequency for the nuclear motion in the potential well $\hbar \omega \approx   57 \ $ meV is much larger than the barrier $\beta$, the nuclei cannot be localized in one of the potential wells.

It is difficult to treat the dynamical JT effect using DFT when many vibrational modes are present as in case of a JT center in the crystal and often the single-mode approximation is made with remarkable success.\cite{Tunneling-PRL2010, Peihong}   
 In the present case, due to the localized nature of the JT-active states (dangling $sp^2$ bond orbitals pointed towards the vacancy), the JT coupling to modes belonging to further neighbor shells is weak (for the second shell, we find $g^\prime \approx g/6$) and since the higher shell stiffness constants are significantly larger than for the first shell for the vacancy center, the single-mode approximation captures the essential physics  in the present case.


The basic features of the collective nuclear-electronic motion may be described by adopting a simple tight-binding approach, familiar from the  electronic structure theory. We write the collective wave function as the linear combination
$
|\Psi\rangle =  c_1\  \phi_1 (R)\  \psi_1^e (R,r) + c_2\  \phi_2 (R)\  \psi_2^e (R,r) + c_3 \  \phi_3 (R)\  \psi_3^e (R,r) ,  
$
where $R (r)$ is the nuclear (electronic) coordinate, $\phi_i(R)$ solves the nuclear Schr\"odinger equation in the vicinity of the  potential minima, 
\begin{equation}
[T_R + V_i (R)] \phi_i(R) = E_0 \phi_i(R),
\end{equation}
and $\psi_i^e (R,r)$ satisfies the electronic Schr\"odinger equation for the fixed nuclear position $R \equiv (Q_1, Q_2)$. The electronic wave function is restricted to the Hilbert space $( |v_1\rangle, |v_2\rangle ) $
and has the form Eq. (\ref{Psi-electronic}) for a given nuclear coordinate $R$. 
Thus the energy eigenstates assume the Born-Oppenheimer form 
$|\Psi (R) \rangle = \Phi_n (R) |\psi_e (R, r) \rangle$,
where $\Phi_n (R )  =    c_1 \phi_1 (R) + c_2 \phi_2 (R)  + c_3 \phi_3 (R)$ is a linear combination of the nuclear orbitals. The eigenstates can then be obtained from the diagonalization of the $3 \times 3$ Hamiltonian 
\begin{equation}
H=
\left(
\begin{array}{ccc}
E_0 & \Gamma  e^{i \phi} & \Gamma  e^{-i \phi}\\
\Gamma e^{-i \phi} &   E_0 & \Gamma  e^{i \phi} \\
\Gamma e^{i \phi} & \Gamma e^{-i \phi}  &   E_0 
\end{array}
\right),
\label{N3X3}
\end{equation}
where the phase factor $e^{i \phi}$ will be discussed momentarily, $E_0$ is the on-site energy, and $\Gamma$ is the nuclear hopping integral in the adiabatic approximation
$ \Gamma = \langle \phi_1 (R) \psi^e (R,r) | \Delta V (R) |   \phi_2 (R) \psi^e (R,r) \rangle  \approx - \Delta V \times F $. In obtaining the last result,
the normalization $ \langle \psi^e (R,r) | \psi^e (R,r) \rangle = 1$ has been used,  $F = \int \phi_1^* (R) \phi_2 (R) d^3 R $ is the Frank-Condon factor, and the deviation of the lower APS potential from the well potential,  $\Delta V (R)  = V_- (R) - V_i (R)$,
has been approximated by its value $-\Delta V$ at the saddle point (marked by a cross in the bottom panel in Fig (\ref{LAPW})), since that's where most of the contribution to the integral comes from.

The magnitude of the nuclear hopping $\Gamma$ may be estimated by assuming a one-dimensional motion of the nuclei in the azimuthal direction, along the circle of radius $\rho_0$ and by computing the quantities $\Delta V$ and $F$. The 1D motion is reasonable since by expanding the adiabatic potential $V_-$ around the potential minima, the spring constant for azimuthal motion is found to be $K' = 9 G$, which  is about half of the spring constant $K$ for radial motion. This corresponds to a phonon frequency of  $ \hbar \omega \approx 58$ meV for radial motion and $\approx 34$ meV for the azimuthal motion. The latter is of the same order of magnitude as  the tunneling barrier of $19$ meV, which again indicates strong tunneling between the three minima.
Now, taking the nuclear wave functions as the 1D simple harmonic oscillator wave function localized at the potential minima: $\phi (x) = (\pi l^2)^{-1/4} \exp \ [- x^2/ (2 l^2)]$, where $ l = \hbar /  \sqrt{M K'}$ and $x$ is the length along the azimuthal direction, the Frank-Condon factor becomes simply the overlap integral between two displaced harmonic oscillator wave functions, with the result: $F = 2^{-1/2} \exp \ [-a^2/ (4 l^2)]$, where $a = 2\pi \rho_0/3$ is the distance between two minima along the circle. Meanwhile, the potential difference between the minimum and the saddle point can be found to be $\Delta V =  \rho_0^2 \times ( \pi^2 K' / 18 - 2G )$.
Plugging in the numerical values, we find $ F \simeq 0.13$ and $\Delta V \simeq 0.035$ eV, so that the hopping integral $ \Gamma \approx \Delta V \times F =  -37 $ cm$^{-1}$.

Finally, in addition to the hopping integral, the adiabatic motion of the electron results in a  fictitious magnetic field seen by the nuclei
 with the  vector potential\cite{Mead}
\begin{equation}
\vec {A} = - \frac{\hbar}{q} \ {\rm Im} \langle \psi_e (R, r) | \vec \nabla_R  \psi_e (R, r) \rangle, 
\label{VecPot}
\end{equation}
which adds a phase factor to the hopping amplitude in the Hamiltonian (\ref{N3X3}).
The modified hopping in the presence of the magnetic field, from point $a$ to $b$, 
is given by  the expression\cite{Feynman}
\begin{equation}
\Gamma  = \Gamma _ {\vec A = 0} \times \exp \big[ \  \frac{iq}{\hbar} \int_a^b \vec{A} \cdot d\vec {s} \ \big]  .
\label{hopping-mag}
\end{equation}
It immediately follows from Eqs. (\ref{Psi-electronic}) and (\ref{VecPot}) that $ \vec {A} =  -2^{-1}\hbar q^{-1} \hat e_\phi$, so that the phase  factor in Eq. (\ref{hopping-mag}) is simply
$e^{i \phi} = e^{i \pi /3}$. This phase factor is very important as without this, the symmetry of the ground state is incorrectly predicted.  Diagonalization of the Hamiltonian Eq. (\ref{N3X3}) with the correct  phase factor yields a doubly-degenerate nuclear ground-state  with energy $\Gamma$, with the singly-degenerate excited state at energy $-2\Gamma$, so that the energy separation is  $ 3 |\Gamma| = 111 $ cm$^{-1}$.

%
\begin{figure}[!htb]
\includegraphics[angle=0,width=6.5cm]{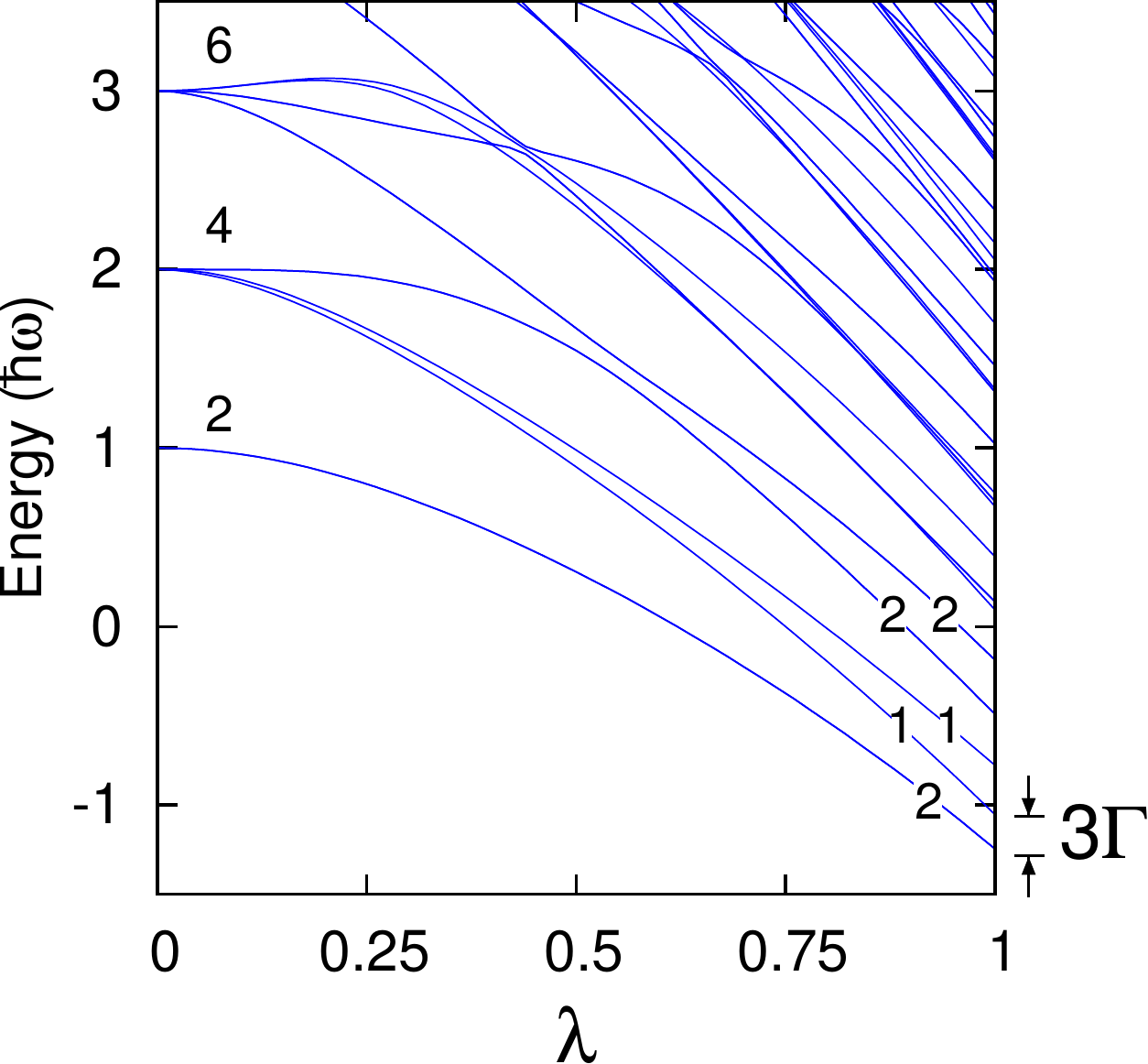} 
\caption{(Color online) Eigenvalues obtained by diagonalization of Eq. (\ref{hamiltonian}) using the basis set Eq. (\ref{expansion}) as a function of the scaled coupling strengths $\lambda g$ and $\lambda G$.  Numbers inside the figure indicates the degeneracies. For $\lambda = 0$, eigenstates of the two-dimensional harmonic oscillator are reproduced.
%
}
\label{Eigenvalue-Curve}
\end{figure}
%

This crude but conceptually rich tight-binding result may be compared to the exact, brute-force  diagonalization of the full    
  Hamiltonian  Eq. (\ref{hamiltonian}) 
by expanding the combined nuclear-electronic wave function $| \Psi  \rangle$ in a complete basis set\cite{Bersuker2006, Allen} 
\begin{eqnarray}
\label{basis-functions}
| \Psi  \rangle  &=& \sum_{n = 0}^N  \sum_{m = 0}^{N-n}   \left [  A_{nm} \frac{(c_1^\dagger)^n}{\sqrt {n!} } 
\frac{(c_2^\dagger)^m}{\sqrt {m!} } |0 \rangle |v_1\rangle      \right.   
 \nonumber \\
&+&    \left.  B_{nm} \frac{(c_1^\dagger)^n}{\sqrt {n!} }  
 \frac{(c_2^\dagger)^m}{\sqrt {m!} } |0 \rangle |v_2\rangle  \right ],
 \label{expansion}
\end{eqnarray}
where  $c_1^\dagger , c_2^\dagger$ create harmonic oscillator states along the $Q_1 , Q_2$ axes centered at the origin and $A_{nm}$ and $B_{nm}$ are the expansion coefficients.  This procedure requires no additional consideration of a fictitious magnetic field and also
yields the full solutions in addition to the lowest three states obtained from the tight-binding theory.
The results are shown in Fig. (\ref{Eigenvalue-Curve}). 
 The magnitude of the tunneling splitting $ 3 |\Gamma| = 86$ cm$^{-1}$  
compares very well with the  tight-binding result.

The large value of the tunneling splitting as compared to the strain splitting, typical value of which\cite{Ham-book-1972} is $\delta \sim $ 10 cm$^{-1}$, results in the delocalization of the nuclear wave function. If the reverse were true, then the nuclei would be more or less stuck in one or the other potential well due to the removal of the degeneracy of the three APS minima by the local strain caused by the invariable presence of defects. This would therefore lead to a static distortion of the nuclear framework resulting in the static JT effect. For the dynamical JT effect, the tunneling splitting must be strong enough to overcome the strain splitting, so that the nuclei can tunnel between all APS minima, which is the case for graphene.

\begin{figure}[!htb]
\includegraphics[width=8.5cm] {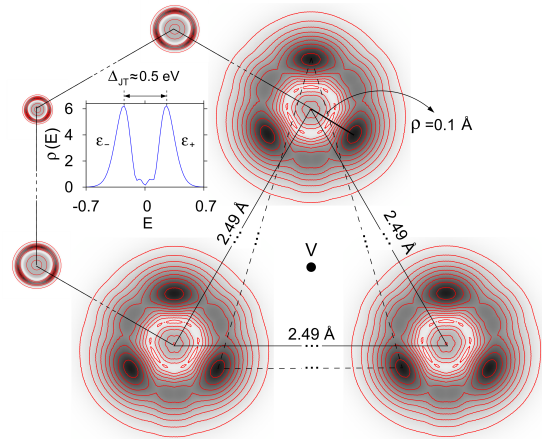}
\caption{(Color online) Nuclear probability density $|\Psi_N (r)|^2$ showing the symmetric distortion of the carbon atoms from the ideal position of an equilateral triangle (solid line). The nuclei move in a correlated  manner so that the most probable configuration is one of the three isosceles triangles (dashed lines) corresponding to the three minima of the APS. The nuclear motion of the nearby atoms show much smaller deviation from their equilibrium positions. {\it Inset} shows  a significant broadening, computed within the adiabatic approximation, of the JT active electron states due to the spread of the nuclear wave function. 
}
\label{Fig4}
\end{figure}

 
 The nuclear probability density in real space $|\Psi_N (r)|^2$ can be computed from the corresponding quantity in the configuration space
 \begin{equation}
  |\Psi_N (Q_1, Q_2)|^2 = \sum_{nm} (|A_{nm}|^2 + |B_{nm}|^2)  |\phi_n(Q_1)|^2   |\phi_m(Q_2)|^2,  
 \end{equation}
 where $\phi_n$ is the n$^{th}$ harmonic oscillator eigenfunction. The calculated $|\Psi_N (r)|^2$ is shown in Fig. (\ref{Fig4}), which indicates a significant spread of the nuclear wave function of the carbon triangle, about $0.1$ \AA\ from the equilibrium positions. We note that this is not washed out by the lattice thermal vibrations, which causes the nuclear vibrational amplitude, estimated from the expression $\frac{1}{2}  K Q^2  = \frac{3}{2} k_B T $ to be about 0.05 \AA\ at room temperature.
 
The spread of the  nuclear wave function broadens the energy of the JT split electronic states as well, so that they are not sharp $\delta$-function states any longer. 
In the adiabatic approximation, the electronic density-of-states is given by
$\rho (E) = \sum_{Q_1 Q_2} |\Psi_N (Q_1, Q_2)|^2 \times [ \delta (E - \varepsilon_- (Q_1, Q_2)) +
\delta (E - \varepsilon_+ (Q_1, Q_2))  $, where $\varepsilon_\pm$ denote the two JT split states as in the expression (\ref{APES}) without the elastic energy term. The results are shown in the inset of Fig. (\ref{Fig4}), which predicts a rather large width, of the order of 0.15 eV, due to the JT effect. Thus these states should appear as rather broad states in scanning tunneling experiments. 
In contrast to this, the broadening of the midgap $V\pi$ state is expected to be rather small. In fact, it is exactly zero if only the nearest-neighbor hopping is retained.\cite{castro-zero-mode} 
This is borne out by the less than 5 meV width of the midgap state, seen in the scanning tunneling  experiments.\cite{Ugeda}  
 

In conclusion, we showed that the substitutional vacancy in graphene forms a dynamical JT center owing to a weak potential barrier for tunneling between the three minima in the adiabatic potential surface. 
The doubly-degenerate nuclear ground state with the tunneling splitting of about 86 cm$^{-1}$ originates from the combined nuclear-electronic motion, which may be cast in terms of a Berry phase acquired due to a fictitious magnetic field experienced by the nuclei caused by the adiabatic motion of the electrons.
The splitting should be observable in the electron paramagnetic resonance and two-photon resonance scattering experiments, which have been used to study the JT effects in the triatomic molecules. The quantum mechanical spread of the nuclear wave function is predicted to lead to a significant broadening of the JT split dangling bond states. Recently, it has been proposed\cite{Entanglement} that the entanglement between the nuclear and electronic motion in a dynamical JT system may be exploited in quantum computation, leading to the possibility of yet another novel application for graphene.

This work was supported by the U. S. Department of Energy through Grant No.
DE-FG02-00ER45818.

* Permanent Address: Institute of Nuclear Sciences, Vin\v{c}a, University of Belgrade, P. O. Box 522, 11001 Belgrade, Serbia

$^\dagger$ Permanent Address: Department of Physics, Indian Institute of Technology Madras, Chennai 600036, India

\end{document}